\documentclass[pra,twocolumn,english,reprint, longbibliography, superscriptaddress, breaklinks=true, showkeys, showpacs=false, nofootinbib]{revtex4-2}

\usepackage[T1]{fontenc}
\usepackage[utf8]{inputenc}
\usepackage{color}
\usepackage{babel}
\usepackage{amsmath}
\usepackage{amssymb}
\usepackage{subfigure}
\usepackage{graphicx}
\usepackage{physics} 
\usepackage{dsfont}
\usepackage{hyperref}
\usepackage{svg}
\usepackage{ulem}
\usepackage{cancel}

\begin{document}

\title{Speedup of thermodynamic entropy production via quantum dynamical criticality}

\author{Andesson B. Nascimento}
\affiliation{QPequi Group, Institute of Physics, Federal University of Goi\'as, Goi\^ania, Goi\'as, 74.690-900, Brazil}

\author{Lucas C. C\'eleri}
\email{lucas@qpequi.com}
\affiliation{QPequi Group, Institute of Physics, Federal University of Goi\'as, Goi\^ania, Goi\'as, 74.690-900, Brazil}

\begin{abstract}
The thermodynamics of quantum phase transitions has long been a rich area of research, providing numerous insights and enhancing our understanding of this important phenomenon. This theoretical framework has been well-developed specially because quantum phase transitions occur at equilibrium. However, its dynamical counterpart, known as dynamical quantum phase transitions (DQPTs), takes place out-of-equilibrium, where conventional thermodynamic tools are inapplicable. In this work, we make progress in this area by connecting dynamical criticality to thermodynamics through a geometric perspective on entropy production. Our findings, along with other recent developments, suggest that dynamical criticality can lead the system to highly complex dynamics, indicating a possible pathway to thermalization. 
\end{abstract}

\maketitle

\section{Introduction}
\label{secI}

Unlike classical phase transitions, a quantum phase transition occurs at absolute zero temperature, driven by quantum fluctuations that cause a sudden change in the phase of the quantum system. At non-zero temperatures, both quantum and thermal fluctuations are present, making the distinction between the two types of critical phenomena less clear~\cite{Sachdev2011}.

Recently, a new category of quantum critical phenomena has been proposed, known as dynamical quantum phase transitions (DQPT)~\cite{Heyl2013}. These transitions occur dynamically over time rather than at equilibrium~\cite{Heyl2013,Heyl2018,Heyl2019}. Contrary to equilibrium cases that focus on asymptotic behavior, DQPTs are concerned with intermediate time scales, where time is the crucial parameter driving the critical phenomena.

DQPTs can be characterized in two distinct ways~\cite{DPT-IandII}. The first, known as DPT-I, involves an order parameter whose time average remains finite in one phase and vanishes in the other, similar to what is observed in equilibrium transitions~\cite{Sciolla2013,Smacchia2015,Halimeh2017,Zhang2017,Chen2020,Bento2024}. The second form, DPT-II, is based on a dynamic analog of free energy per particle, called the Loschmidt echo~\cite{Heyl2014,Heyl2015,Vajna2015,Bhattacharya2017,Heyl2017,Flaschner2018,Jurcevic2017,Goes2020}. These two forms are independent and in principle there is no direct correspondence between equilibrium and dynamical critical phenomena. Interesting connections between these types of dynamical criticality were discussed in Refs.~\cite{Corps2022,Corps2023}.

A key characteristic of DQPT is that it occurs out-of-equilibrium, complicating the study of its thermodynamics. This emerging field uses various tools from non-equilibrium physics to understand dynamical criticality. In Ref.~\cite{Goes2020}, the authors investigated entropy production in quantum phase space using Wehrl entropy, finding that entropy increases over time on average in systems presenting dynamical criticality. Ref.~\cite{Bento2024} took a different approach, demonstrating that the complexity of dynamics, closely related to entropy production, increases much faster when the system exhibits dynamical criticality. These findings suggest the feasibility of developing a thermodynamic theory for dynamical criticality.

In the present work, we advance this direction by considering a geometric perspective on entropy production in the context of dynamical criticality. Entropy production is a fundamental quantity in equilbrium physics, as its positivity indicates irreversibility~\cite{EPgeo,EPirrev,EPcorr,EPopensys} (see Ref.\cite{EPrev} for a review). Links between geometry and thermodynamics are not new. In the classical realm, we can mention the seminal works by Ruppeiner~\cite{Ruppeinere1995}, who employed differential geometry to build a thermodynamic theory for equilibrium. For quantum systems, a geometric perspective can shed new light on several phenomena because it describes quantum mechanics and the classical phase space on equal footing~\cite{Strocchi1966,Heslot1985}. In particular, this approach helps us understanding the thermodynamics of quantum phase transitions. It is worth noting that a geometric theory for both dynamical quantum phase transitions~\cite{Lang2018} and quantum thermodynamics~\cite{Anza2022} has been recently discussed, although from distinct perspectives.

Of special interest to the present work, we mention the relation between the Bures angle and the thermodynamic entropy production~\cite{Deffner2010,Deffner2013}. We explore such connection in order to develop a thermodynamic description of type II dynamical quantum phase transition. This is achieved by linking entropy production and the Loschmidt echo by means of the Bures angle, which characterizes this type of criticality. Specifically, we confirm that thermodynamic entropy is produced much faster for critical systems. In other words, dynamical criticality speeds up entropy production. 
This is significant because the geometric measure of entropy production  provides a more direct and quantifiable link to thermodynamic entropy, unlike Wehrl entropy~\cite{Goes2020} or complexity~\cite{Bento2024}, which only suggest this connection. It is important to observer here that, although we employed a specific model to perform the numerical studies, the main result is model independent, since it is rooted into the main features of the dynamical quantum phase transition. Moreover, since there are connections between both types of DQPTs, we expect our results to qualitatively hold for DQPT-I as well.

The paper is structures as follows. We present a brief discussion on DQPT in Sec.~\ref{secII} in terms of Loschmidt echo, in which we define the quantities of interest for our study. An overview of the geometric approach to entropy production is presented in Sec.~\ref{secIII}. We then introduce the relationship between the entropy production and the Bures angle, which we use in Sec.~\ref{secIV} to compute a lower bound on the entropy production and its time-average for a specific model. Finally, in Sec.~\ref{secV} we present our final discussions.

\section{Dynamical quantum phase transition}
\label{secII}

In this section, we provide a brief description of dynamical quantum phase transitions. Specifically, we consider the so called type II DQPT~\cite{DPT-IandII} which is characterized through the nonanalytical behavior of the rate function (see Eq.~\eqref{ratefunc} bellow) at the critical times. Our intention here is to establish the notation and present the basic equations. This will be important in the next section. We point the reader to Refs.~\cite{Heyl2018,Heyl2019} for deeper discussions.

Let us consider a closed quantum system that can be externally controlled by means of some parameter $\xi$. At the beginning of the protocol ($\xi \equiv \xi_{i}$), the system is prepared in the state $\ket{\psi_0}$ that belongs to the ground state manifold of the initial Hamiltonian $H_{i} \equiv H(\xi_{i})$. A quench is then applied to the system, changing its Hamiltonian according to $H_{i} \rightarrow H_{f} \equiv H(\xi_{f})$ and the system is let to evolve under $H_{f}$. 

The central quantity in the theory of DQPT is the Loschmidt amplitude $\mathcal{G}_{t}$, which is defined by the return amplitude 
\begin{equation}
    \mathcal{G}_{t} = \braket{\psi_0}{\psi_t},
\end{equation}
with $\ket{\psi_t} = e^{-iH_f t}\ket{\psi_0}$ being the evolved state at time $t$.

The identification of the DQPT is based on a formal analogy between $\mathcal{G}$ and the bounded partition function~\cite{Heyl2018,Heyl2019}. In this sense, we can define the rate function in terms of the Loschmidt echo $L_{e} = |\mathcal{G}_t|^2$ as
\begin{equation}
    \lambda(t) = - \lim_{N\xrightarrow{} \infty} \frac{1}{N} \log L_e(t),
    \label{ratefunc}
\end{equation}
with $N$ being the system size. Such a quantity plays here the role of a thermodynamic potential, like the free-energy in the theory of equilibrium quantum phase transitions. Shortly, we extend $\mathcal{G}_t$ to $\mathcal{G}_z$, where $z = t + i\tau$ is the complex time. In the thermodynamic limit, the Fisher zeros of $\mathcal{G}$ cross the real time axis, indicating the occurrence of a quantum phase transition in the same way that happens in the equilibrium case. Therefore, we can understand $\lambda$ as a partition function, bounded by the ground state~\cite{Heyl2013}.

Although Eq.~\eqref{ratefunc} is defined in terms of a limiting process, for a finite system we can write
\begin{equation}
    L_e(t) = e^{-N\lambda(t)}.
    \label{ecorate}
\end{equation}

This approach is well justified both experimentally and theoretically. In experiments, it is impossible to deal with infinite systems, but the signatures of DQPTs can be observed with high precision in finite systems, as demonstrated in \cite{Zunkovic2018, Jurcevic2017, Heyl2014}. In this context, the analysis of DQPTs in systems with $\mathbb{Z}_2$ symmetry breaking and ground state degeneracy allows us to capture critical features even in finite systems. The ground state degeneracy is particularly relevant since, in broken symmetry phases, as observed in the Ising \cite{Zunkovic2018, Jurcevic2017} and XXZ models \cite{Heyl2014}, the rate function $\lambda(t)$ can be expressed in terms of the contributions from different symmetry sectors, $\lambda_{\eta}(t)$, which represent the probability of returning to the ground state. 

In systems with degenerate ground states, the probability of returning to the ground state manifold can be written as
\begin{equation}
P(t) = \sum_{\eta} P_{\eta}(t),
\end{equation}
where $P_{\eta}(t) = |\langle \eta | \psi_0(t) \rangle|^2$ is the probability of the system returning to one of the ground states $\ket{\eta}$, and the sum is taken over all possible degeneracies of the system.

For large systems $N \gg 1$, each of these probabilities follows a large deviation scaling, $P_{\eta}(t) = e^{-N\lambda_{\eta}(t)}$ \cite{Heyl2014}, where $\lambda_{\eta}(t)$ is the rate function associated with the return probability $P_{\eta}(t)$. As $N$ increases, one of the probabilities $P_{\eta}(t)$ dominates \cite{Jurcevic2017}, which is equivalent to having the smallest rate function become effective in this regime. 

In the thermodynamic limit, the different functions $\lambda_{\eta}(t)$ converge to a global rate function $\lambda(t)$, which can be defined as
\begin{equation}
    \lambda(t) = \min_{\eta} \lambda_{\eta}(t),
\end{equation}
which implies that only one of the probabilities associated with the ground state degeneracies dominates the system's behavior, allowing the return probability to be written in a form similar to the eq. (\ref{ecorate}).

The dominance shifts from one probability to another at a time, producing a non-analyticity in the effective rate function $\lambda(t)$ \cite{Heyl2014}. A DQPT is characterized by the appearance of this non-analytic point in $\lambda(t)$, with the time at which this occurs being the critical time, $t_c$.

Although this formulation is rigorously valid for infinite systems, it was shown in \cite{Heyl2014} that DQPTs can be detected in finite systems, as the temporal non-analyticities associated with the Loschmidt echo remain robust even for moderately sized systems.

This robustness allows DQPTs to be observed even in numerical simulations of finite systems. In \cite{Zunkovic2018}, DQPTs were analyzed in an Ising chain with long-range interactions, where the rate function $\lambda(t)$ exhibited non-analyticities across different interaction ranges, with the dynamical transitions clearly visible in finite-sized systems. Similarly, in \cite{Heyl2014}, transitions in the XXZ model showed comparable non-analyticities. Both models present a degenerated ground state and $\mathbb{Z}_2$ symmetry.

Thus, by following this approach, we assume that although the thermodynamic limit provides a rigorous description of DQPTs, the essential features of these transitions already emerge in finite systems.

This analysis can be extended to the case of mixed states~\cite{Bhattacharya2017,Heyl2017}. In this case, the purification is employed in order to define the generalized Loschmidt echo as $\mathcal{G} = \Tr{\rho_{0}^{p} U_{t}}$, where $U_{t}$ is the unitary evolution operator while $\rho_{0}^{p} = \ket{w(0)}\bra{w(0)}$ stands for the purified density matrix. Here, $\ket{w(0)}=\sum_{i} \sqrt{p_{i}}\ket{\psi_{i}^{s}(0)}\otimes\ket{\psi_{i}^{a}}$ is the normalized initial state acting on the joint Hilbert space (system $s$ plus ancilla $a$) $\mathcal{H}_{p} = \mathcal{H}_s\otimes \mathcal{H}_a$. The state of the system is obtained through the partial trace over the ancilla, $\rho = \Tr_{a}{\rho_{0}^{p}}$.

This will be important for us here since we are interested in thermodynamics, thus making the connection to thermal states unavoidable. However, we can think about it in an alternative way. We start in a pure state and, at the end, we just need to take into account the entropy generated by this preparation, which can be accounted for through the distance between the pure state and the reference thermal state. Since we are interested in the dynamics of entropy production, this fixed amount of initial entropy will not be important. This fact will become clear in the next section.

The goal of the next section is to build a relation between the rate function and the entropy production by using a geometric formulation of thermodynamics.

\section{Geometric approach to entropy production}
\label{secIII}

Geometric statistical distances, which are employed to study the evolution of physical states, can be understood as measures of distances between probability distributions~\cite{StaDist}. In the context of quantum mechanics, considering two pure states $\ket{\psi_1}$ and $\ket{\psi_2}$, a useful quantity is the Wootters' distance~\cite{Wootters}
\begin{equation}
    d_w(\ket{\psi_1},\ket{\psi_2}) = \arccos{|\bra{\psi_1}\ket{\psi_2}|},
    \label{WoottAng}
\end{equation}
which measures the angle in state space between the considered states. Additionally, it is the only monotone Riemannian metric, up to a constant factor, invariant under all unitary transformations~\cite{Wootters}, thus ensuring that the metric is preserved under temporal evolution~\cite{WoottersAngle}.

A generalization to the space of density matrices was proposed in Ref.~\cite{Bures}. The distance between two infinitesimally close states $\rho$ and $\rho + \dd\rho$ is the Bures metric
\begin{equation}
    D_{B}^{2}(\rho + \dd \rho, \rho) = \frac{\Tr(\dd \rho G)}{2},
\end{equation}
with the operator $G$ being defined by the equation $\{\rho, G\} = \dd \rho$~\cite{Deffner2013}. In terms of the eigenbasis of $\rho = \sum p_{i}\dyad{i}$, we have
\begin{equation}
    D_{B}^{2}(\rho + \dd \rho, \rho) =\sum_{i,j} \frac{|\bra{i} \dd\rho \ket{j}|^2}{2(p_i + p_j)}
\end{equation}
Similarly to Wootters' distance, the Bures metric can be considered a natural metric on the space of density matrices and, in the limit case in which $\rho$ is pure, the Bures metric reduces to Wootters' statistical distance~\cite{StaDist}. Also, we can interpret Bures distance as the angle between two mixed states in the space of density matrices.

Considering two density operators, $\rho_1$ and $\rho_2$, the Bures distance $\mathcal{L}(\rho_1, \rho_2)$, also known as Bures angle, can be defined in terms of Uhlmann Fidelity~\cite{Uhlmann} $F(\rho_1, \rho_2) = [\Tr(\sqrt{\rho_1}\rho_2\sqrt{\rho_1})^{\frac{1}{2}}]^2$, which is symmetric, non-negative and invariant under unitary transformations. The Bures angle can be written as~\cite{Deffner2013,Deffner2010} 
\begin{equation}
    \mathcal{L}(\rho_1,\rho_2) = \arccos{\sqrt{F(\rho_1,\rho_2)}}.
    \label{BuresFidelity}
\end{equation}
It is important to note here that, even though fidelity is not a metric due to its failure to satisfy the triangle inequality, the Bures angle defined in Eq.~\eqref{BuresFidelity} is~\cite{Nielsen}.

Now, in order to put this discussion in the context of DQPT, we consider the Bures angle between the pure ground state $\rho_0$ of the initial system Hamiltonian and the evolved state $\rho_t$, under the quenched Hamiltonian. The fidelity can be expressed as $F(\rho_0,\rho_t) = \Tr(\rho_0\rho_t) = |\braket{\psi_0}{\psi_t}|^2$ and the Bures angle takes the simple form
\begin{equation}
    \mathcal{L}(\rho_0,\rho_t) = \arccos{|\bra{\psi_0}\ket{\psi_t}|},
    \label{Bures}
\end{equation}
which, due to Eq.~\eqref{ecorate}, can be written as
\begin{equation}
 \mathcal{L}(\rho_0,\rho_t) = \arccos\sqrt{L_{e}(t)} = \arccos{e^{-\frac{N}{2}\lambda(t)}}.
\label{BuresRate}
\end{equation}
Note that we have disregarded the thermodynamic limit here, which presents no problem as long as the quantum quench does not induce superextensive energy fluctuations in the system~\cite{Heyl2017} (see also the discussion in the previous section). Therefore, the rate function is directly linked to the Bures angle. We can see that, at the critical point where the Loschmidt echo vanishes, the distance between the initial and the evolved states becomes maximum. Although the Loschmidt echo oscillates in time, its amplitude decays to zero much faster when the quench crosses the critical point --compared to the case where there is no dynamical criticality. This is a general behaviour of dynamical critical systems. This indicates that the system tends to thermalize in a state that is orthogonal to the ground state manifold, thus reaching the maximum of the Bures angle. This will be related to the maximum of the entropy production in the next section.

Again, such discussion can be extended to the case of mixed states, as we discussed  previously. However, this will not be necessary since we do not need to take into account the entropy produced by the initial preparation of the pure state, which is always the fixed ground state in our case. For instance, we can start the system in a thermal state and then prepare it in the ground state, by a projection of simply by cooling it down. According to Ref.~\cite{Deffner2011}, the entropy relative to this process contributes with a constant that would simply rescale the values of entropy production. Since our focus is on the time evolution of entropy production during the dynamical process, this entropy do not affect the qualitative behavior of the results.

We are now ready to discuss the entropy production from this geometric perspective and to establish its connections to the dynamical critical behavior.

\subsection{Entropy Production}
\label{subsecA}

Entropy production $\langle \Sigma \rangle$ is a fundamental concept in non-equilibrium thermodynamic that is deeply linked to irreversibility. It is associated with non-equilibrium fluxes, like heat, for instance, due to existence of a thermodynamic force, like a temperature gradient~\cite{Deffner2011}.

For nonequilibrium phenomena, there is always a non-zero entropy production, $\langle \Sigma \rangle \geq 0$, which holds as equality only for reversible processes~\cite{Deffner2010, Deffner2013}. However, the Clausius inequality does not account for how far from equilibrium the process operates.

A generalization of the Clausius inequality was derived in~\cite{Deffner2010}, where a lower bound on entropy production $\langle \Sigma \rangle$ is given in terms of the Bures angle. The proposal precisely quantify the relationship between entropy production and the irreversibility of the process, i.e., how far the process occurred from equilibrium, using geometric tools. The result is process-dependent and valid for arbitrary nonequilibrium driving, extending beyond the linear response regime.

The same authors proposed in ~\cite{Deffner2013} a more refined result to lower bound to entropy production, by incorporating higher-order terms in the series expansion, leading to a sharper and more accurate lower bound, thus allowing for a better quantification of entropy production, especially in systems far from equilibrium, including those subjected to quench dynamics.

\begin{figure}[h]
    \begin{center}
    \includegraphics[width=0.5\textwidth]{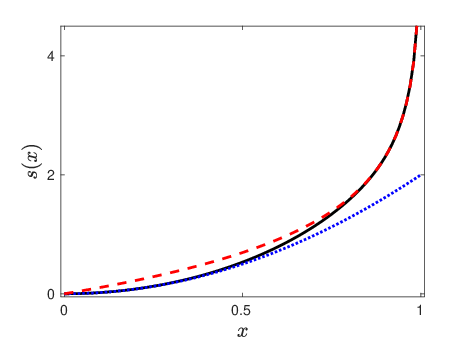}
    \end{center}
    \caption{The lower bound on entropy production. The solid (black) line shows the exact function $s(x)$, while the dashed (red) and the dotted (blue) show the upper and lower bounds on $s(x)$, respectively. The variables in the axis are dimensionless.}
    \label{s(x)}
\end{figure}

By considering that the system is initially in a thermodynamic equilibrium state such lower bound can be put in terms of the Bures angle as ~\cite{Deffner2013}
\begin{equation}
    \langle \Sigma \rangle  \geqslant s\left(\frac{2}{\pi} \mathcal{L}(\rho_{\tau},\rho_{\tau}^{eq}) \right),
    \label{lowerbound}
\end{equation}
where 
\begin{equation}
s(x) = \min_{x<r<1} S((r-x,1-r+x)\vert\vert (r,1-r))
\label{f_s(x)}
\end{equation}
for $0 \leq x < 1$. In this equation, $(p,1-p)$ represents a two-dimensional probability distribution while $S(p\vert\vert q)$ is the relative entropy (or Kullback-Leibler divergence) between the distributions $p$ and $q$. From the series expansion we can find~\cite{Eisert} 
\begin{equation}
    2x^{2} \leq s(x) \leq -\ln(1-x).
    \label{bounds}
\end{equation}
Such bounds are illustrated in Fig.~\ref{s(x)}.

The lower bound in Eq.~\eqref{bounds} corresponds to the result of Ref.~\cite{Deffner2010}. The Bures angle offers a natural way to link entropy production to the system's distance from equilibrium. This provides a clear understanding that entropy production increases as the system is driven out of equilibrium, making the Bures angle an effective tool for quantifying irreversibility.

A comment regarding the purity of our initial state is in order here. The entropy produced in the system by some process can be written as $S(\rho_{t}\vert\vert\rho_{\beta})$, where $\rho_{t}$ is the evolved state of the system due to the considered process while $\rho_{\beta}$ is the reference final equilibrium state. In other words, the entropy is quantified by how well we can distinguish between the actual state and the equilibrium one. This results is obtained by considering that the system is initially in the thermal state defined by the initial Hamiltonian and the inverse temperature $\beta$. If the initial state of the system is not an equilibrium one, we must consider the entropy generated by the process of measurement, that takes the system out of equilibrium. In our case, this is given by $S(\rho_{0}\vert\vert\rho_{\beta})$, with $\rho_{0}$ being the ground state of the system. This is just a constant that will not matter to Eq.~\eqref{lowerbound} and, thus, can be safely neglected in the analysis that follows.  

Now, since $\mathcal{L}$ oscillates as a function of time ---it is a function of the Loschmidt echo--- we expect that the time average of both the entropy production and of the bound are more informative than the quantities themselves. This does not represent an actual violation of the second law since we are dealing with an out of equilibrium system, so positivity of entropy production is not strictly expected. It is interesting to observe that although $\mathcal{L}$ oscillates in time, when the critical point is crossed, its amplitude decays very fast, indicating a complete distinct behavior in both phases. That is the origin of the results we present here.

Before entering in the calculations of entropy production, we can drawn some conclusions from these bounds. When the system is very close to a critical time, $\mathcal{L}$ approaches $\pi/2$ and, therefore, the upper bound of $s$ diverges, thus implying that entropy production can grow without limits. Actually, we expect a divergent behavior of the entropy production at the critical point, as usually occurs in critical phenomena. On the other hand, when the system is far away from the critical point, the rate function becomes very small, and the lower bound on the entropy production, which is quadratic in $\mathcal{L}$, is very close to zero, implying that we can have a very low amount of entropy being produced in the system. Thus, we see that the entropy production can indeed sign a DQPT. Note that although we expect an oscillatory behavior of the entropy production, since we are discussing the second law of thermodynamics, we also expect an increasing of this quantity on average, towards a maximum value. It is important to observe here that these are general conclusions that are model independent since all systems presenting dynamical criticality will display the same qualitative features.

\section{Entropy production across a DQPT}
\label{secIV}

In order to investigate the dynamics of entropy production we consider the paradigmatic Lipkin-Meshkov-Glick (LMG) model~\cite{LMG1,LMG2,LMG3}, which describes a fully connected $N$ spin-$1/2$ system. The LMG model is defined by the Hamiltonian
\begin{equation}
    \hat{H} = -2h\hat{J}_z - \frac{g}{j}\left(\hat{J}_{x}^{2} + \gamma \hat{J}_{y}^{2}\right)
    \label{H_lmg}
\end{equation}
where, $j = N/2$, $\hat{J}_l = \sum_i \sigma_{l}^{i}/2$ are collective spin operators, with $\sigma_{l}^{i}$ being the Pauli matrix in direction $l$ associated to spin $i$. $h$, $\gamma$ and $g$ are coupling constants. From here on we choose $\gamma = 1/2$ and $g=1$, thus setting the energy scale of the problem. Since we are setting $\hbar=1$ this choice of $g$ makes time dimensionless.

The dynamical critical point of this model is $h^{d}_{c} = (h_0 + g)/2$ for quenches starting at $h=h_{0}$~\cite{pappalardi}. Here we consider, for simplicity, quenches starting at $h_{0} = 0$, thus implying $h^{d}_{c} = 1/2$.

The Hamiltonian of the LMG model possesses $\mathbb{Z}_2$ parity symmetry \cite{Mzaouali2021}. For $h_0 = 0$, the initial state of the system is in the spontaneously broken symmetry phase, meaning that even though the Hamiltonian (\ref{H_lmg}) is invariant under spin-flip transformation, the initial state of the system is not. When we perform a quench that crosses the dynamical critical point of the model, the system evolves into a state where parity symmetry is dynamically restored. Therefore, the system has two distinct dynamical phases: a symmetric phase and a broken-symmetry phase. Furthermore, the ground state of the model is doubly degenerate. These two characteristics make the LMG model share the same conditions as the XXZ and Ising models discussed in section \ref{secII}, thus allowing us to use the approach introduced there.

\begin{figure}
  \begin{center}
        \includegraphics[width=0.5\textwidth]{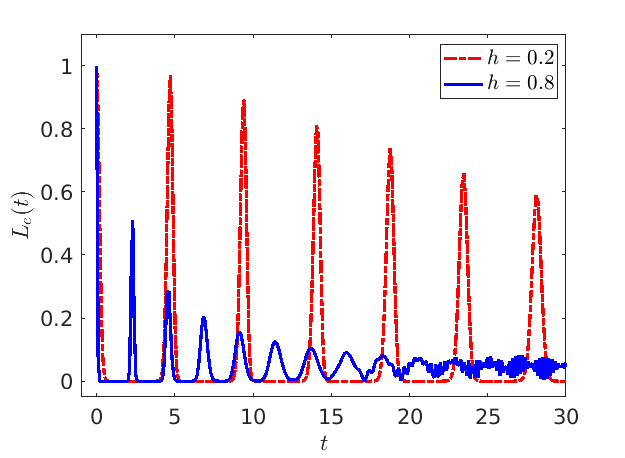}
        \includegraphics[width=0.5\textwidth]{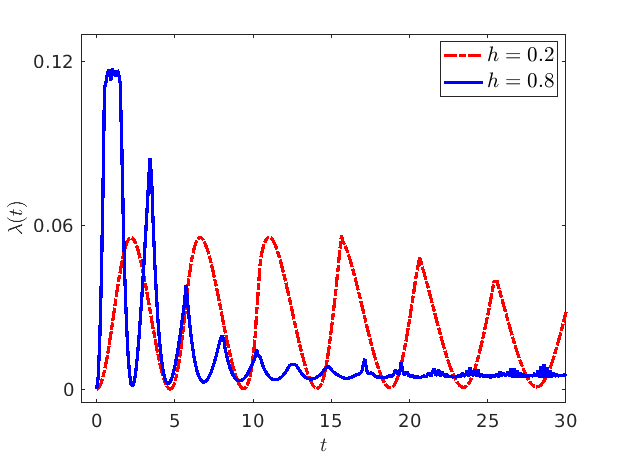}
  \end{center}  
    \caption{The top figure illustrates the behavior of the Loschmidt echo for the LMG model under two distinct quenches: one crossing the dynamical critical point (blue solid lines) and the other not (red dashed lines). The bottom panel displays the rate function for the same quenches. In both plots, we consider $j=300$ and $\gamma = 1/2$. The dynamical critical point is $h_{c}^{d} = 1/2$.}
    \label{Echo_rate}
\end{figure}

The LMG model, given by Eq.~\eqref{H_lmg}, shows a type II DQPT, as we can see in Fig.~\ref{Echo_rate}, where the behavior of the Loschmidt echo (upper panel) and the rate function (botom panel) are presented. We can clearly observe the non-analytic behavior of the rate function when the quench crosses the dynamical critical point (blue solid line), characterizing the DQPT in the system.

\begin{figure}[h!]
    \begin{center}
    \includegraphics[width=0.45\textwidth]{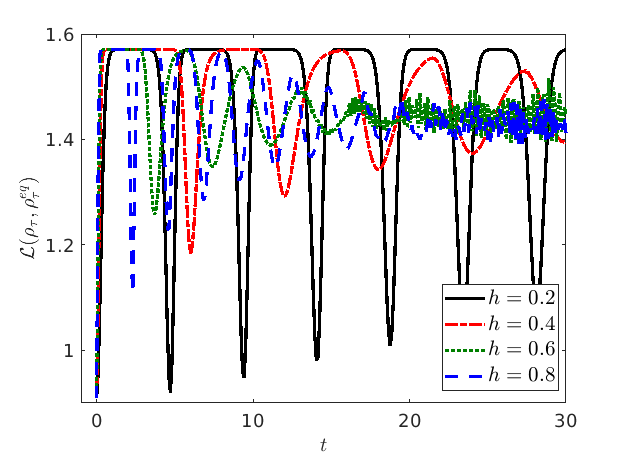}
    \includegraphics[width=0.45\textwidth]{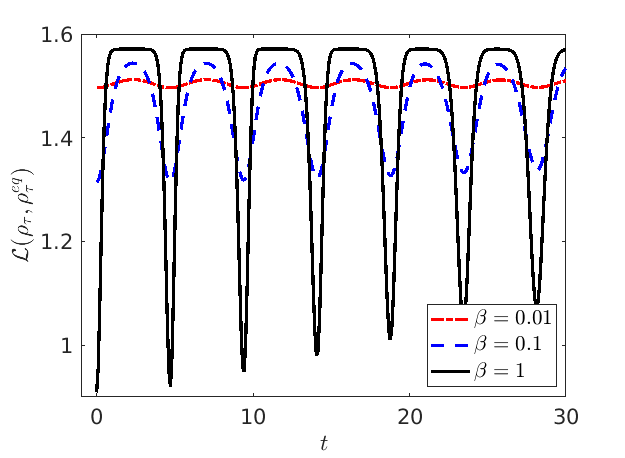}
    \includegraphics[width=0.45\textwidth]{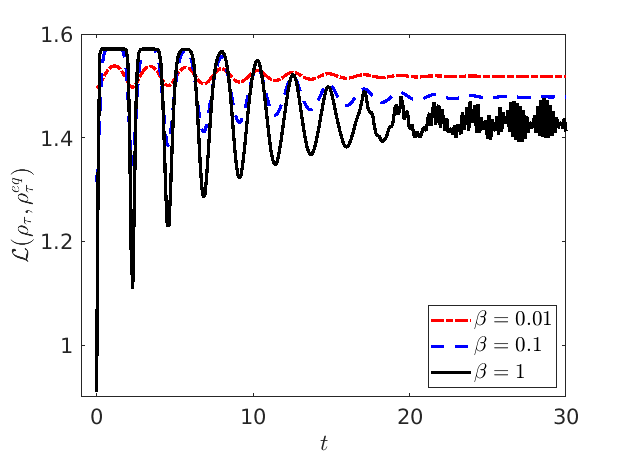}
    \end{center}
    \caption{The top panel shows the dynamics of the Bures angle for four distinct quenches, $h=(0.2,0.4,0.6,0.8)$. The middle panel shows the effect of the temperature when the quench does not cross the critical point ($h=0.2$), while the bottom one shows the same behavior when we cross the critical point ($h=0.8$). For all the plots we consider $j=300$.}
    \label{BA}
\end{figure}

Let us start by analyzing the behavior of the Bures angle, which determines the bound on the entropy production. In order to make contact to thermodynamics, we consider the initial thermal state 
\begin{equation}
    \rho_0 = \frac{e^{-\beta H_0}}{Z_0},
\end{equation}
where $Z_0 = \mathrm{Tr}(e^{-\beta H_0})$ is the partition function. The system Hamiltonian is quenched from $h_0 = 0$ to the final value $h$ and it is then allowed to evolve under the quenched Hamiltonian. In this setup, Fig.~\ref{BA} shows the behavior of the Bures angle for distinct quenches and different temperatures.

It is clear from the figure that the dynamical behavior of $\mathcal{L}$ is very different for both dynamical phases. While in both cases it oscillates in time with decreasing amplitude, when the quench cross the critical point, the rate at which the decay occurs is much larger compared with the case in which we do not crosses the critical point. Since this quantity lower bounds the entropy production, this result indicates that the bound on the thermodynamic entropy saturates much faster when the system presents dynamical criticality. In other words, entropy will be produced in the system due to the complex dynamics of a many-body system, but dynamical criticality highly increases the speed at which this happens. This is necessary in order to keep the second law of thermodynamic valid, since $s(x)$ is a monotonic function of its argument.

Such dynamics aligns with our discussion in the previous section. As illustrated in Fig.~\ref{Echo_rate}, when the quench crosses the critical point, the rate function $\lambda(t)$ (and also the Loschmidt echo) exhibits a decaying oscillation amplitude during the time evolution that is much faster than what happens in the case of no dynamical criticality. Since this is a general behavior of systems displaying DQPT, the results presented here are model-independent.

Regarding the effect of temperature, we can see that when it increases ($\beta$ decreases), the bound tends to a constant, thus implying that thermal fluctuations are much more important than the quantum fluctuations behind the dynamical criticality. This shows that our results are consistent with what is expected from thermodynamic considerations. By increasing $\beta$ (decreasing temperature) we are approaching the ground state of the system, thus recovering the contribution coming solely from the dynamical criticality, which is induced by quantum fluctuations.

\begin{figure}[h!]
    \centering
        \includegraphics[width=\linewidth]{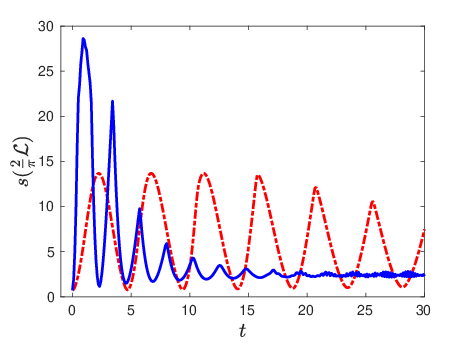} \\
    \caption{Lower bound on the entropy production. The solid (blue) line shows the case $h=0.8$, (crossing the critical point) while the dashed (red) one displays the case $h=0.2$ (not crossing the critical point). We considered $j=300$ and $\beta = 1$.}
    \label{EPRate}
\end{figure}

In order to be more precise, we show in Fig.~\ref{EPRate} the dynamical behavior of the lower bound $s(\frac{2}{\pi}\mathcal{L})$ for the entropy production. The distinct dynamics is clear from the figure. The bound tends to stabilize with a much greater rate when the quench crosses the critical point. As anticipated, this indicates that when the system undergoes a DQPT, the lower bound of entropy production tends to stabilize rapidly. This suggests that the system reaches the equilibrium in a much shorter time when the quench crosses the dynamical critical point.

\begin{figure}[h!]
    \centering
    \includegraphics[width=0.5\textwidth]{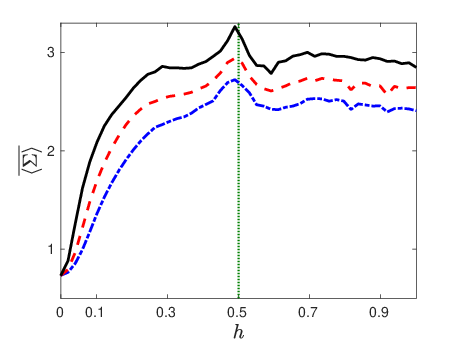}
    \caption{The figure illustrates the $h$-dependent behavior of the time-averaged entropy production for three distinct size spin chains, $j=100$ (blue doted-dashed line), $j=200$ (red dashed line) and $j=500$ (black solid line). We considered $T=10^3$ to perform the calculation of the time average presented in the graph.}
    \label{TAEP}
\end{figure}

As mentioned before, since we are dealing with a non-equilibrium system, it is interesting to investigate the time average of the entropy production 
\begin{equation}
    \overline{\langle \Sigma \rangle} = \lim_{T\rightarrow\infty}\frac{1}{T}\int_{0}^{T} \dd t \langle \Sigma \rangle.
\end{equation} 
This is illustrated in Fig.~\ref{TAEP}  as function of the quench parameter $h$. We can identify two main features of entropy production in this figure. First, we can see that entropy production can indeed signs the DQPT given its distinct behaviours on both phases. Secondly, it is evident that the entropy reaches a plateau, which indicates a stabilization much faster when the quench crosses the critical point. 

Actually, such features are expected to hold in all critical models since, from our discussion, this behavior is a consequence of the dynamics of the rate function, which is qualitatively the same for all models.

\section{Conclusion}
\label{secV}

Thermodynamics and phase transitions are deeply interconnected topics in both classical and quantum physics. In this work, we advance the development of a thermodynamic theory for dynamical quantum phase transitions. These critical phenomena resemble quantum phase transitions but are fundamentally different, as they occur during the system's evolution over time rather than at equilibrium. We achieve this by adopting a geometric perspective on entropy production. This approach enables us to establish a relationship between the Loschmidt echo, which characterizes dynamical criticality, and entropy production through the Bures angle, which is a geometric measure of the distance between quantum states in the state space.

Our main result can be summarized as follows. First, we observe that the thermodynamic entropy production (in its geometric form) signals the dynamical quantum phase transition. Unlike in the equilibrium case, the dynamics of entropy production is the crucial factor here. The rate of entropy production in different phases of the quantum system is markedly different, providing clear evidence of the complexity of the dynamics when the system exhibits dynamical criticality. Secondly, dynamical criticality accelerates entropy production. Generally, entropy is produced in a quantum many-body system due to the complex nature of its dynamics. However, our findings show that when the system exhibits dynamical critical behavior, entropy is generated at a significantly greater rate. 

These results have several implications for understanding this type of phenomenon. One notable implication is that rapid entropy production saturation raises questions about the thermalization of closed quantum systems. The model we consider here, the LMG model, is integrable. In Ref.~\cite{Bento2024}, it was discovered that dynamical criticality drives the system towards an equal probability distribution in the energy eigenbasis. Together with the findings of the present work, this suggests that such critical phenomena may drive the system towards equilibrium. However, a deeper analysis is necessary. The eigenstate thermalization hypothesis~\cite{Deutsch2018} might provide further insights for this line of investigation.

Another interesting line of investigation concerns the relationship between the generation of quantum coherences with respect to the energy eigenbasis, as well as entanglement, and entropy production. This is in the same spirit as Refs.~\cite{Oliveira2024,Celeri2024}, where these quantities were directly linked, albeit in a completely different context.

Finally, considering all of this, as well as other results linking DQPT and, for instance, information scrambling~\cite{Heyl2018b,Goes2020}, it is not too speculative to ask if dynamical critical systems share some features with quantum chaotic systems regarding their ergodic properties.

\begin{acknowledgments}
This work was supported by the National Institute for the Science and Technology of Quantum Information (INCT-IQ), Grant No. 465469/2014-0, by the National Council for Scientific and Technological Development (CNPq), Grants No 308065/2022-0, and by Coordination of Superior Level Staff Improvement (CAPES).
\end{acknowledgments}


\end{document}